\documentclass[sigconf]{acmart}

\usepackage{cite} 
\usepackage{natbib}
\setcitestyle{numbers,sort&compress}

\usepackage{graphicx}
\usepackage{multirow}
\usepackage{booktabs}
\usepackage{amsmath}
\usepackage{url}
\usepackage{enumitem}
\usepackage{float}
\usepackage{subcaption}
\usepackage{makecell}
\usepackage[normalem]{ulem}
\usepackage{microtype}
\usepackage{tabularx}
\usepackage{xspace}
\newcommand{\Semantic}{Semantic}
\usepackage{algorithm}
\usepackage{algorithmic}

\newcommand{\modelname}{SymCERE\xspace}

\title{SymCERE: Symmetric Contrastive Learning for Robust Review-Enhanced Recommendation}

\author{Toyotaro Suzumura}
\email{suzumura@acm.org}
\affiliation{%
  \institution{The University of Tokyo / Rakuten Group, Inc.}
  \state{Tokyo}
  \country{Japan}
}

\author{Hisashi Ikari}
\email{hisashi@ikari.io}
\affiliation{%
  \institution{Independent Researcher}
  \city{Tokyo}
  \country{Japan}}

\author{Hiroki Kanezashi}
\email{hkanezashi@acm.org}
\orcid{0000-0002-8329-6235}
\affiliation{%
  \institution{The University of Tokyo / Rakuten Group, Inc.}
  \state{Tokyo}
  \country{Japan}
}

\author{Md Mostafizur Rahman}
\email{mdmostafizu.a.rahman@rakuten.com}
\affiliation{%
  \institution{Rakuten Group, Inc.}
  \city{Tokyo}
  \country{Japan}}

\author{Yu Hirate}
\email{yu.hirate@rakuten.com}
\affiliation{%
  \institution{Rakuten Group, Inc.}
  \city{Tokyo}
  \country{Japan}}

\begin{document}

\begin{abstract}
Modern recommendation systems fuse user behavior graphs and review texts but often encounter a "Fusion Gap" caused by False Negatives, Popularity Bias, and Signal Ambiguity. We propose SymCERE (Symmetric NCE), a contrastive learning framework bridging this gap via structural geometric alignment. First, we introduce a symmetric NCE loss that leverages full interaction history to exclude false negatives. Second, we integrate L2 normalization to structurally neutralize popularity bias. Experiments on 15 datasets (e-commerce, local reviews, travel) demonstrate that SymCERE outperforms strong baselines, improving NDCG@10 by up to 43.6\%. Notably, we validate this on raw reviews, addressing significant noise. Analysis reveals "Semantic Anchoring," where the model aligns on objective vocabulary (e.g., "OEM," "gasket") rather than generic sentiment. This indicates effective alignment stems from extracting factual attributes, offering a path toward robust, interpretable systems. The code is available at \url{https://anonymous.4open.science/r/ReviewGNN-2E1E}.
\end{abstract}

\maketitle

\section{Introduction}

Recent advances in recommender systems focus on multi-modal approaches that fuse user behavior, typically captured by Graph Neural Networks (GNNs), with side information from modalities like review texts, processed by Large Language Models (LLMs). While this fusion promises richer user and item representations \citep{Zhao2024DynLLM, Luo2024MoLAR}, its practical implementation faces a triad of interconnected challenges—a phenomenon we identify as the "Fusion Gap". This gap represents the disconnect between the high-quality semantic potential of LLMs and the noisy geometric reality of collaborative filtering spaces.

The first challenge contributing to this gap is the false negative problem, a known limitation in standard contrastive learning (CL) objectives like InfoNCE. By treating all non-positive pairs as negatives, standard methods inherently penalize "false negatives"—\textbf{valid positive items (whether known interactions or latent preferences) erroneously treated as negatives}. This creates "intra-class repulsion" \citep{Tsai2022Incremental, Huynh2022Boosting}, pushing semantically related items apart in the embedding space and degrading the coherent structure learned by the LLM.

\begin{figure}[t]
    \centering
    \includegraphics[width=\linewidth]{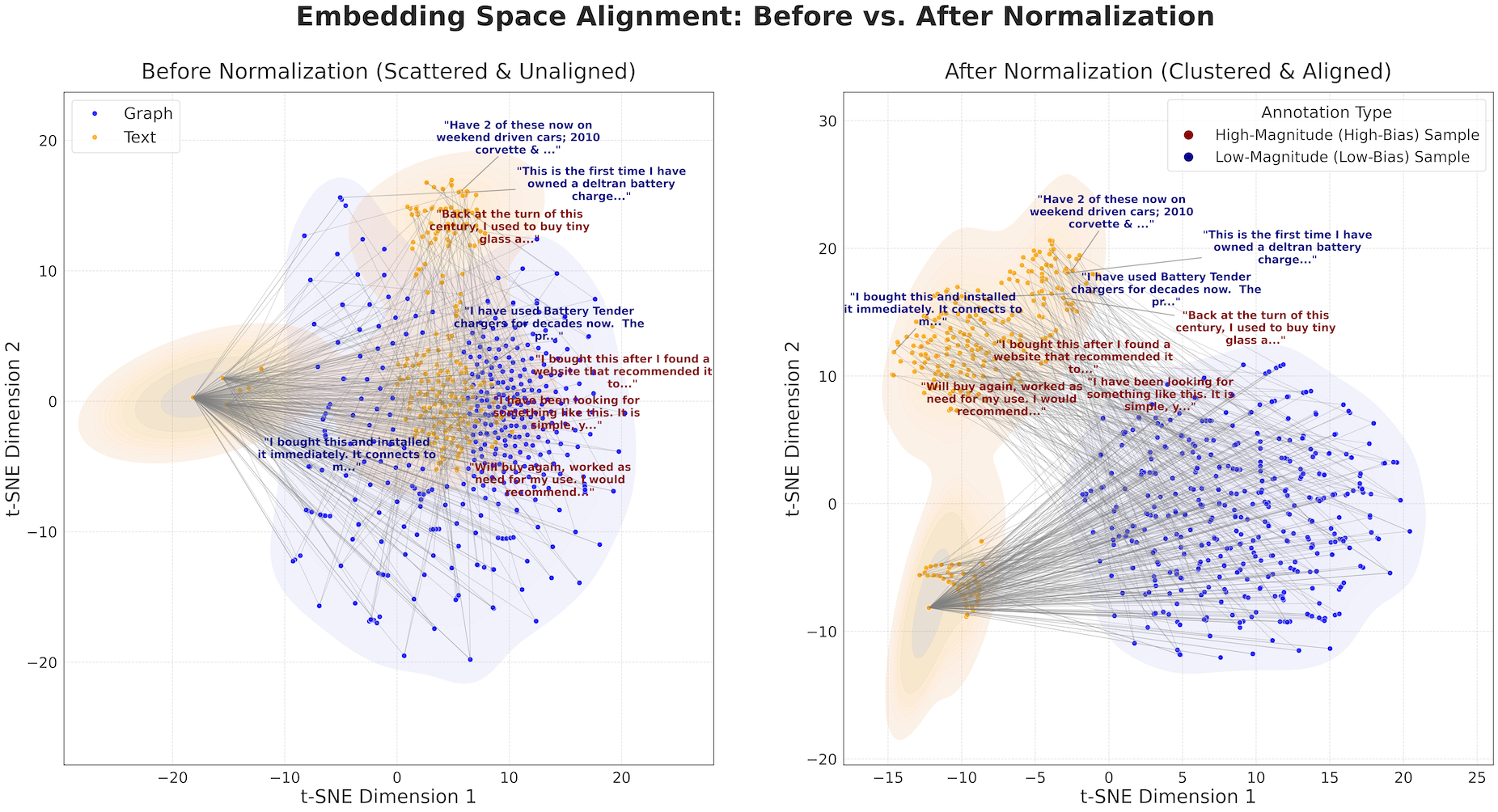} 
    \caption{Visualizing the Fusion Gap and its Resolution. Left: Before normalization, the embedding space is dominated by magnitude artifacts (popularity bias), causing a geometric and semantic disconnect between Graph (blue) and Text (orange) modalities. Right: SymCERE structurally aligns these spaces on a unit hypersphere, enabling coherent clustering based on semantic preference rather than interaction frequency.}
    \label{fig:fusion_gap_scatter}
\end{figure}

The second challenge is popularity bias, which acts as a geometric artifact. Popular items, due to their high interaction frequency, tend to develop embeddings with a disproportionately large L2-norm (magnitude) \citep{Klimashevskaia2024Survey, Jannach2021Survey}. We hypothesize that this magnitude acts as a confounding noise factor, reflecting popularity rather than a user's intrinsic preference. When fusing modalities, this magnitude-driven noise can overshadow the delicate directional signals from the text modality (as visualized in Figure \ref{fig:fusion_gap_scatter}, Left).

The third challenge is signal ambiguity, which complicates search-to-recommendation bridging. Unlike clean product metadata (used in many MMRec tasks), a user review contains valuable semantic signals about product attributes (e.g., "This OEM part fits perfectly.") entangled with subjective, noisy sentiment (e.g., "I love it!") \citep{Wu2019Survey_review, Bauman2017Recommending}. A model that cannot geometrically distinguish these signals may learn spurious correlations from generic sentiment instead of factual alignment \citep{Cheng2024Empowering}.

While various methods have been proposed to tackle these issues individually—using denoising modules \citep{Yao2023Denoising}, causal inference for debiasing \citep{Zhang2021Causal, Zheng2021Disentangling}, or heuristic false-negative correction \citep{Tsai2022Incremental}—we explore a unified geometric approach. We posit that these three challenges can be systematically mitigated by modifying the core contrastive learning objective to respect the geometry of the embedding space.

We propose SymCERE (Symmetric NCE), a unified contrastive learning methodology that addresses this triad to bridge the Fusion Gap. First, to mitigate the false negative problem, it employs a symmetric, denoised contrastive loss based on NCE, which uses the known interaction graph to deterministically remove in-batch false negatives, reducing intra-class repulsion. Second, guided by our "magnitude-as-noise" hypothesis, it projects all embeddings onto a unit hypersphere via L2 normalization. This structurally removes the popularity signal to isolate the preference encoded in the vector's direction. Finally, we demonstrate that a model built on these principles naturally resolves signal ambiguity, learning to anchor its alignment not on superficial sentiment but on objective, information-rich features.

Our main contributions are:
\begin{itemize}[leftmargin=*]
    \item A Unified Geometric Framework (SymCERE): We propose a method that structurally mitigates false negatives, popularity bias, and signal ambiguity. It achieves this by symmetrically applying a denoised NCE loss for cross-modal alignment and integrating geometric debiasing via L2 normalization.
    \item Extensive Empirical Validation: We demonstrate the effectiveness of SymCERE through comprehensive experiments on 15 diverse datasets, achieving significant improvements (e.g., up to 43.6\% in NDCG@10) over strong baselines.
    \item Discovery of Semantic Anchoring: We find, through in-depth LIME analysis, that our model learns to prioritize product-specific terminology (e.g., `OEM', `compatible') over generic sentiment (e.g., `good', `nice'). We term this mechanism `Semantic Anchoring', providing evidence that robust multi-modal alignment is driven by factual product attributes.
\end{itemize}

\section{Related Work}
\label{sec:related_work}
This work lies at the intersection of multi-modal recommendation, Graph Neural Networks (GNNs), and contrastive learning (CL). GNNs like LightGCN \citep{He2020LightGCN} form the backbone of many modern recommenders by modeling the user-item interaction graph. To enhance the learned representations, contrastive learning (CL) has been widely adopted, typically using the InfoNCE loss with augmented data \citep{Wu2021Self}. However, this common approach may overlook the potential of CL to address deeper, interconnected geometric problems.

\subsection{A Triad of Interconnected Challenges}
We frame the primary obstacles in review-enhanced recommendation not as isolated issues requiring bespoke modules, but as an interconnected triad that results in a "Fusion Gap."

\subsubsection{The False Negative Problem}
The standard InfoNCE loss has a key limitation: by treating all non-positive samples as negatives, it inherently penalizes "false negatives"—semantically similar items that are not the anchor's direct positive pair \citep{Huynh2022Boosting, Tsai2022Incremental}. This "intra-class repulsion" negatively affects the embedding space by pushing similar items apart. While heuristic correction methods exist \citep{Tsai2022Incremental}, they can be complex. In recommendation, we have the advantage of a known interaction history, which allows for more deterministic mitigation strategies.

\subsubsection{Popularity Bias as a Geometric Artifact}
Popularity bias, the over-recommendation of popular items, is a known challenge to fairness and diversity \citep{Klimashevskaia2024Survey}. We approach this from a geometric perspective, positing that popularity is often encoded as the magnitude (L2-norm) of an item's embedding vector \citep{Xu2022Neutralizing}. This magnitude acts as a noisy signal that obscures user preferences. Therefore, we treat L2 normalization not just as a technical step, but as a debiasing mechanism that structurally removes the popularity signal to encourage the model to encode preference solely in the vector's direction \citep{Wang2020Understanding}.

\subsubsection{Signal Ambiguity in Multi-Modal Fusion}
Fusing collaborative signals with review texts enriches representations but introduces "signal ambiguity". A review conflates objective product attributes (e.g., "OEM part," "compatible gasket") with subjective sentiment (e.g., "good," "great") \citep{Wu2019Survey_review}. A simple alignment may lead to learning spurious correlations from generic sentiment. We hypothesize that a robustly formulated CL method can resolve this ambiguity by forcing alignment based on stable, objective features, effectively using reviews as "soft metadata."

\subsection{Our Approach: A Unified Solution via Principled CL}
The literature often presents a range of methods that address these issues individually for denoising, debiasing, and multi-modal fusion \citep{Tang2024Robust, Yao2023Denoising, Ma2024Triple}. We diverge from this trend, arguing that modifying the core CL objective itself offers a more direct and unified solution.
To this end, we build upon NCE, a supervised contrastive loss that addresses the limitations of the InfoNCE objective by reformulating it to eliminate intra-class repulsion. By integrating a symmetric NCE loss with geometric debiasing (L2 normalization), SymCERE is designed to simultaneously mitigate false negatives and popularity bias, thereby resolving signal ambiguity.

\section{Methodology}
\begin{figure*}[t]
    \centering
    \includegraphics[width=\textwidth]{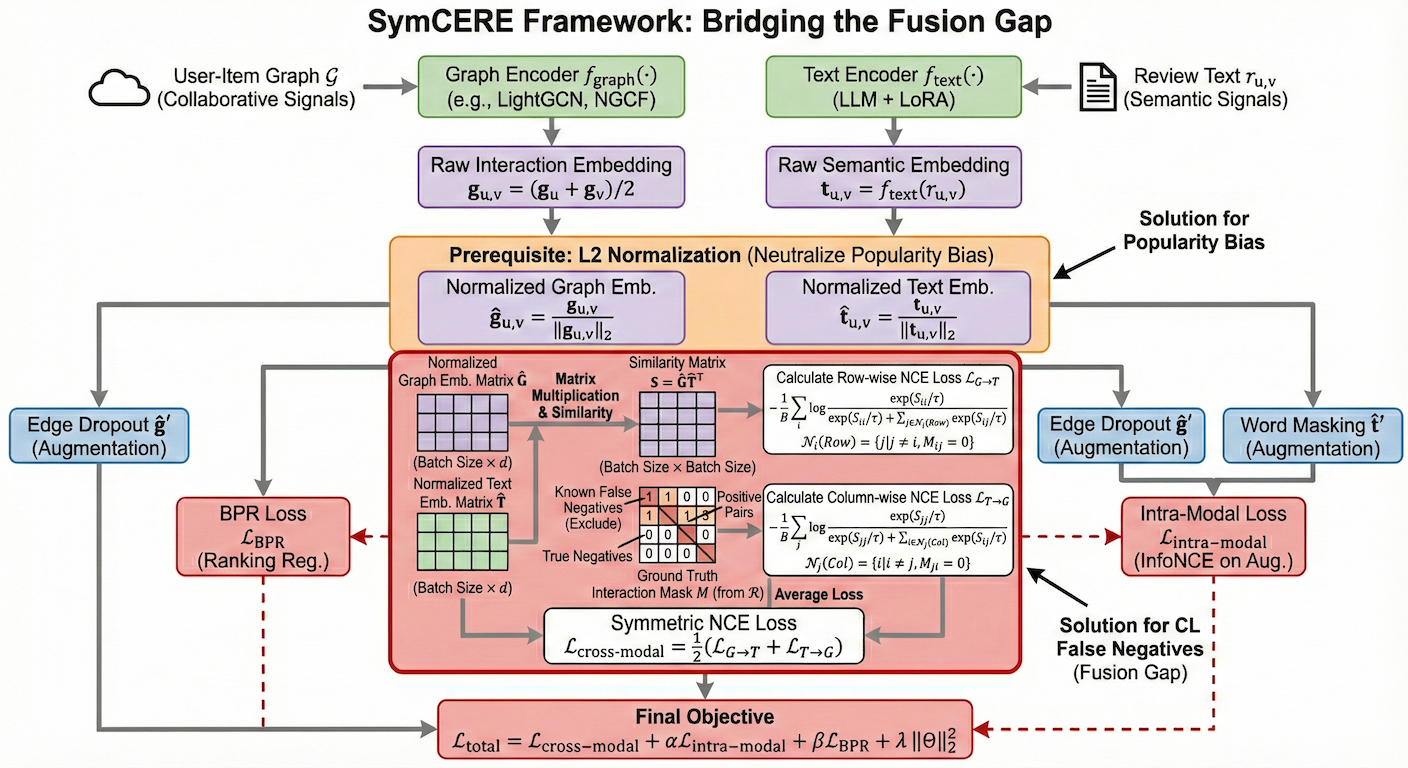}
    \caption{The SymCERE Framework. The model bridges the Fusion Gap through two key geometric mechanisms: (1) L2 Normalization projects embeddings onto a hypersphere to structurally neutralize popularity bias (magnitude noise). (2) A Symmetric NCE Loss utilizes the known interaction mask $M$ to deterministically exclude false negatives (grey cells) from the denominator, preventing intra-class repulsion. This unified objective aligns collaborative signals ($\mathbf{g}$) and semantic signals ($\mathbf{t}$) while preserving ranking order via BPR.}
    \label{fig:symcere_overview}
\end{figure*}

Our proposed framework, \modelname, aligns collaborative signals (graph modality) and semantic information (text modality) into a unified hyperspherical space. The overall architecture is illustrated in Figure \ref{fig:symcere_overview}.

\subsection{Modality Encoders}

\subsubsection{Graph Encoder (Collaborative Signals).}
We employ established GNN architectures as the graph encoder, $f_{\text{graph}}(\cdot)$. In this study, we utilize both LightGCN~\citep{He2020LightGCN} (a linear model) and NGCF~\citep{Wang2019Neural} (a non-linear model) as backbones to evaluate the framework's versatility.

LightGCN refines embeddings $\mathbf{E}^{(0)}$ over $K$ layers via linear propagation. The final representation is a weighted combination of all layers:
\begin{equation}
\mathbf{g}_u = \sum_{k=0}^{K} \alpha_k \mathbf{e}_u^{(k)}, \quad \mathbf{g}_i = \sum_{k=0}^{K} \alpha_k \mathbf{e}_i^{(k)}.
\end{equation}
where $\alpha_k = 1/(K+1)$.
Regardless of the backbone, the interaction $(u,v)$ representation is computed as $\mathbf{g}_{u,v} = (\mathbf{g}_u + \mathbf{g}_v) / 2$. This interaction-centric representation serves as the target for cross-modal alignment.

\subsubsection{Text Encoder (Semantic Signals).}
A pre-trained LLM adapted with LoRA serves as the text encoder, $f_{\text{text}}(\cdot)$. Given review text~$r_{u,v}$ for interaction $(u,v)$, the encoder produces a dense semantic embedding $\mathbf{t}_{u,v} = f_{\text{text}}(r_{u,v})$.
To ensure temporal integrity and prevent leakage, $r_{u,v}$ consists strictly of the review text written by user $u$ for item $v$ at the time of interaction. We treat this text as a source of static, intrinsic product attributes ("Semantic Anchoring") rather than dynamic sentiment, facilitating robust alignment.

\subsection{Unified Contrastive Alignment Framework}
\modelname addresses the geometric distortions caused by popularity bias and contrastive false negatives through a principled fusion objective.

\subsubsection{Prerequisite: L2 Normalization for Stability.}
\label{sec:l2_norm}
To mitigate popularity bias, which often manifests as magnitude artifacts~\citep{Xu2022Neutralizing}, we enforce L2 normalization on all embeddings before loss calculation. This projects representations onto a unit hypersphere, ensuring alignment is based purely on directional similarity:
\begin{equation}
\hat{\mathbf{g}}_{u,v} = \frac{\mathbf{g}_{u,v}}{\|\mathbf{g}_{u,v}\|_2}, \quad \hat{\mathbf{t}}_{u,v} = \frac{\mathbf{t}_{u,v}}{\|\mathbf{t}_{u,v}\|_2}.
\end{equation}

\subsubsection{Multi-Task Contrastive Objectives.}
We combine cross-modal alignment with intra-modal regularization.

\paragraph{Cross-Modal Alignment with Symmetric NCE Loss ($\mathcal{L}_{\text{cross-modal}}$).}
Standard InfoNCE loss suffers from the false negative problem, where unobserved but valid positive items are treated as negatives, causing intra-class repulsion. To resolve this, we propose a Symmetric NCE Loss based on Noise-Contrastive Estimation~\citep{Gutmann2010Noise}. This objective modifies the standard contrastive formulation to deterministically exclude false negatives.

In recommendation, the full interaction history $\mathcal{R}$ is known. For a mini-batch $\mathcal{B}$, we construct a "true negative" set $\mathcal{N}_i$ for an anchor interaction $i=(u_i, v_i)$ by excluding any sample $j$ where the user $u_i$ has a known interaction with item $v_j$:
\begin{equation}
\mathcal{N}_i = \{j \in \mathcal{B} \mid j \neq i \land (u_i, v_j) \notin \mathcal{R}\}.
\end{equation}
This ensures that the denominator only penalizes items that are truly unobserved for the user.
The Graph-to-Text alignment loss ($\mathcal{L}_{G \to T}$) is defined as:

\begin{equation}
\begin{split}
\mathcal{L}_{G \to T} = -\frac{1}{|\mathcal{B}|} \sum_{i \in \mathcal{B}} \Big[ &\frac{\hat{\mathbf{g}}_{u_i,v_i}^\top \hat{\mathbf{t}}_{u_i,v_i}}{\tau} \\
&- \log \Big( \sum_{k \in \{i\} \cup \mathcal{N}_i} \exp \frac{\hat{\mathbf{g}}_{u_i,v_i}^\top \hat{\mathbf{t}}_{u_k,v_k}}{\tau} \Big) \Big].
\end{split}
\end{equation}


Symmetrically, the Text-to-Graph loss ($\mathcal{L}_{T \to G}$) is computed by swapping the roles of anchor and positive, ensuring bidirectional alignment. The final cross-modal loss is:
\begin{equation}
\mathcal{L}_{\text{cross-modal}} = \frac{1}{2} \left( \mathcal{L}_{G \to T} + \mathcal{L}_{T \to G} \right).
\end{equation}

\paragraph{Intra-Modal Enhancement ($\mathcal{L}_{\text{intra-modal}}$).}
To improve the robustness of individual modalities, we employ standard InfoNCE loss on augmented views (edge dropout for graph $\hat{\mathbf{g}}'$, word masking for text $\hat{\mathbf{t}}'$):
\begin{equation}
\mathcal{L}_{\text{intra-modal}} = \mathcal{L}_{\text{InfoNCE}}(\hat{\mathbf{g}}, \hat{\mathbf{g}}') + \mathcal{L}_{\text{InfoNCE}}(\hat{\mathbf{t}}, \hat{\mathbf{t}}').
\end{equation}

\paragraph{Intra-Modal Ranking Regularization.}
We further regularize the graph embedding space using the Bayesian Personalized Ranking (BPR) loss to enforce relative preference ordering:
\begin{equation}
\mathcal{L}_{\text{BPR}} = \sum_{(u,v^+,v^-) \in \mathcal{D}} -\log\sigma(\hat{\mathbf{g}}_u^\top \hat{\mathbf{g}}_{v^+} - \hat{\mathbf{g}}_u^\top \hat{\mathbf{g}}_{v^-}).
\end{equation}

\subsection{Final Objective Function}
The final training objective is a weighted combination of these components:
\begin{equation}
\mathcal{L}_{\text{total}} = \mathcal{L}_{\text{cross-modal}} + \alpha \cdot \mathcal{L}_{\text{intra-modal}} + \beta \cdot \mathcal{L}_{\text{BPR}} + \lambda \|\Theta\|^2_2.
\end{equation}
where $\alpha=0.5$ and $\beta=0.05$ are hyperparameters determined empirically.

\subsection{Theoretical Analysis: Geometric Mechanism of Semantic Anchoring}
\label{sec:theoretical_analysis}

Our empirical results suggest that \modelname exhibits ``Semantic Anchoring,'' prioritizing objective product attributes over subjective sentiment. We provide a geometric proof that this implies the model suppresses components that are \textit{non-discriminative} with respect to the collaborative signal.

\begin{figure}[t]
    \centering
    \includegraphics[width=\columnwidth]{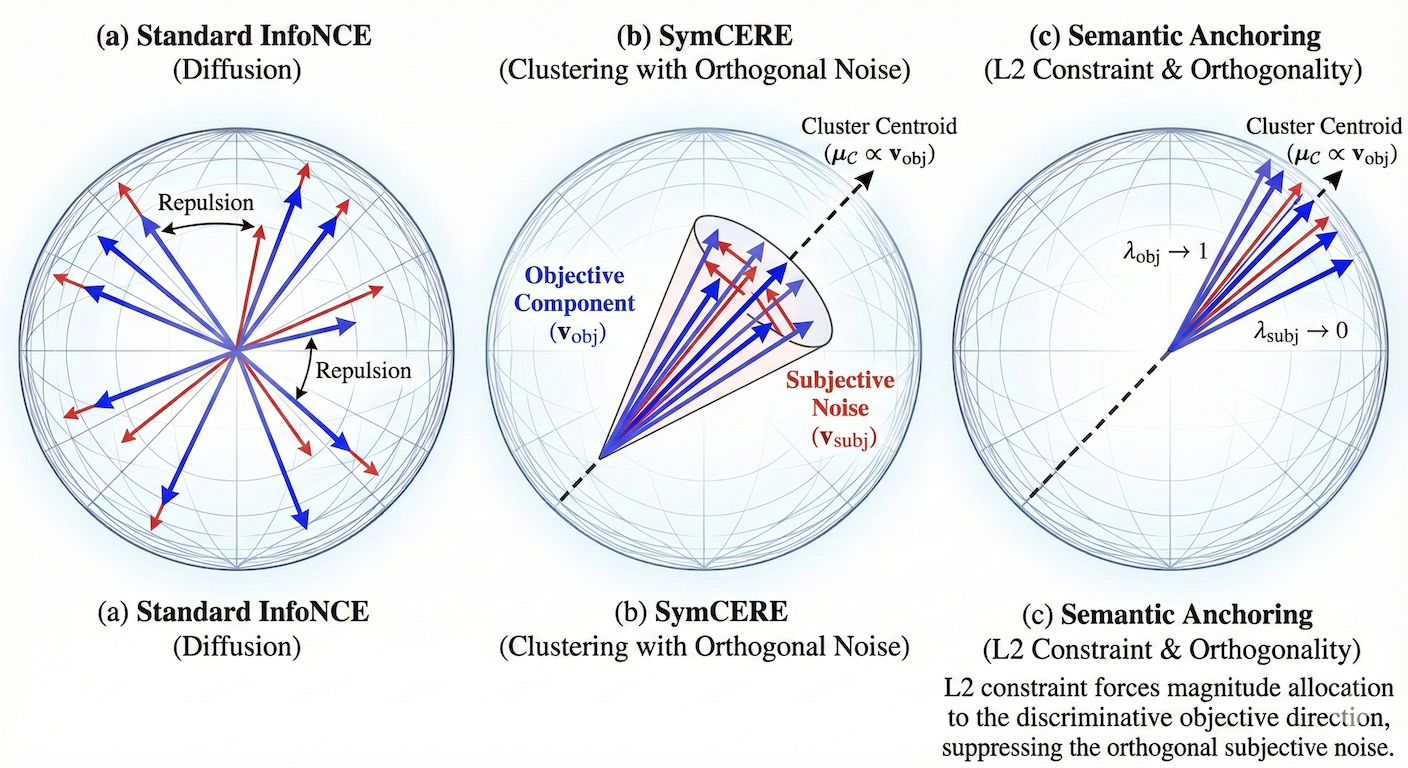}
    \caption{Geometric Mechanism of Semantic Anchoring. (a) Standard InfoNCE induces repulsion between semantically similar items (false negatives), causing diffusion on the hypersphere. This allows subjective noise ($\mathbf{v}_{\text{subj}}$) to persist as there is no structural pressure to align. (b) SymCERE removes these repulsive forces via false negative elimination, allowing items to form tight clusters around a common centroid $\boldsymbol{\mu}_{\mathcal{C}}$ aligned with objective features. (c) \Semantic Anchoring: Under the strict L2 normalization constraint, maximizing alignment with the centroid forces the embedding to allocate its limited magnitude almost entirely to the objective component ($\lambda_{\text{obj}} \to 1$), effectively suppressing subjective noise ($\lambda_{\text{subj}} \to 0$).}
    \label{fig:geometric_mechanism}
\end{figure}

\subsubsection{Vector Decomposition on the Hypersphere}
Consider a semantic embedding vector $\mathbf{t}$ on the unit hypersphere $\mathbb{S}^{d-1}$. We decompose $\mathbf{t}$ into a \textit{discriminative} objective component ($\mathbf{v}_{\text{obj}}$) and a \textit{common-mode} subjective component ($\mathbf{v}_{\text{subj}}$):
\begin{equation}
\mathbf{t} = \lambda_{\text{obj}} \mathbf{v}_{\text{obj}} + \lambda_{\text{subj}} \mathbf{v}_{\text{subj}}, \quad \text{s.t.} \quad \|\mathbf{t}\|_2 = 1.
\end{equation}

\subsubsection{Discriminative Orthogonality under Graph Constraints}
Strictly removing false negatives via $\mathcal{N}_i$ forces the model to align the anchor with a set of semantically related items $\mathcal{C}$ (implicit cluster), maximizing the alignment with the cluster centroid $\boldsymbol{\mu}_{\mathcal{C}}$.
Crucially, the collaborative graph links items based on specific functional substitutability (e.g., specific car parts), not generic sentiment. Thus, the centroid $\boldsymbol{\mu}_{\mathcal{C}}$ is defined by shared objective features:
\begin{equation}
\boldsymbol{\mu}_{\mathcal{C}} \propto \mathbf{v}_{\text{obj}}.
\end{equation}
Conversely, generic sentiment (e.g., "good") appears across diverse, unrelated clusters. While not necessarily zero-mean, it offers no contrastive power to distinguish cluster $\mathcal{C}$ from others. In the local tangent space of the optimal cluster, the subjective component is effectively orthogonal to the discriminative direction:
\begin{equation}
\mathbf{v}_{\text{subj}} \perp \boldsymbol{\mu}_{\mathcal{C}}.
\end{equation}

\subsubsection{Signal Maximization under L2 Constraint}
To maximize the inner product $\mathbf{t}^\top \boldsymbol{\mu}_{\mathcal{C}}$ under the constraint $\lambda_{\text{obj}}^2 + \lambda_{\text{subj}}^2 \approx 1$, the model must allocate magnitude solely to the component parallel to $\boldsymbol{\mu}_{\mathcal{C}}$. Allocating magnitude to $\mathbf{v}_{\text{subj}}$ wastes the limited L2 budget on a non-discriminative dimension:
\begin{equation}
\lim_{\text{training} \to \infty} \lambda_{\text{subj}} \to 0, \quad \lim_{\text{training} \to \infty} \lambda_{\text{obj}} \to 1.
\end{equation}
This proves that ``Semantic Anchoring'' is a structural consequence: L2 normalization forces the model to discard features—even strong ones like sentiment—that do not align with the fine-grained geometric structure imposed by the graph.

\section{Experiments}
We conducted comprehensive experiments to answer:
\begin{itemize}[leftmargin=*]
    \item RQ1: Does SymCERE outperform state-of-the-art GNN-based and self-supervised recommendation baselines?
    \item RQ2: How do false negative mitigation and geometric debiasing contribute to performance? (Ablation studies)
    \item RQ3: What is the semantic mechanism driving the alignment? (Case study)
\end{itemize}

\subsection{Experimental Setup}

\subsubsection{Datasets}
We curated 15 datasets from three sources: 13 Amazon Reviews categories \citep{McAuley2015Image}, Yelp \citep{yelp_dataset}, and a semi-public Travel dataset \citep{rakuten2020travel} to test generalizability across languages and domains. We applied 5-core filtering (except for Travel) to ensure data quality. Statistics are summarized in Table \ref{tab:dataset_stats}.

\begin{table}[htbp]
\centering
\caption{Basic statistics of the datasets used in our experiments.}
\label{tab:dataset_stats}
\begin{tabular}{@{}lrrr@{}}
\toprule
\textbf{Dataset Category} & \textbf{Users} & \textbf{Items} & \textbf{Reviews} \\
\midrule
Clothing Shoes and Jewelry & 1,786 & 4,291 & 13,676 \\
CDs and Vinyl & 2,078 & 3,888 & 24,604 \\
Tools and Home Improvement & 1,290 & 4,571 & 18,456 \\
Sports and Outdoors & 1,372 & 4,190 & 18,660 \\
Beauty and Personal Care & 1,405 & 5,066 & 23,870 \\
Automotive & 1,187 & 2,833 & 12,106 \\
Toys and Games & 1,045 & 4,152 & 17,568 \\
Pet Supplies & 1,114 & 4,241 & 20,722 \\
Health and Household & 820 & 3,679 & 14,442 \\
Video Games & 870 & 2,812 & 17,572 \\
Cell Phones and Accessories & 401 & 1,943 & 6,872 \\
Patio Lawn and Garden & 282 & 1,373 & 3,944 \\
Office Products & 256 & 1,287 & 3,828 \\
Yelp & 802 & 1,274 & 11,879 \\
Travel & 1,275 & 9,809 & 20,365 \\
\bottomrule
\end{tabular}%
\end{table}

\subsubsection{Evaluation Protocol}
For each user, we split their interaction history chronologically into training, validation sets with a ratio of 80:20. This ensures that the model is evaluated on its ability to predict future interactions based on past behavior.
We adopt a top-K item ranking evaluation protocol. For each user in the test set, the model is tasked with ranking the ground-truth item against all other items in the catalogue that the user has not interacted with in the training set. This "all-ranking" protocol provides a comprehensive and unbiased evaluation.
We evaluate model performance using two standard top-K ranking metrics \citep{Jarvelin2002Cumulated, Herlocker2004Evaluating}: HR@10 (Hit Ratio at 10) and NDCG@10 (Normalized Discounted Cumulative Gain at 10).

\subsubsection{Baselines and Comparison Scope}
We compare our model against a suite of competitive baselines covering three categories. First, for collaborative filtering, NeuMF \citep{He2017Neural}, NGCF \citep{Wang2019Neural}, and LightGCN \citep{He2020LightGCN} represent the evolution from neural CF to graph-based methods. Second, regarding graph contrastive learning, SGL \citep{Wu2021Self} serves as a baseline for self-supervised learning on user-item graphs. Third, in the category of multi-modal recommendation, we include SMORE \citep{Zhou2023Freedom} as a state-of-the-art multi-modal contrastive learning baseline. Comparing against SMORE specifically allows us to test the "Fusion Gap" hypothesis—whether simple aggregation fails under noisy review data.

It is important to note that many recent Multi-Modal Recommendation (MMRec) works focus on Product Metadata (clean descriptions/images) (e.g., Amazon Product Data). In contrast, our work targets the more challenging and noisy domain of User Reviews (Amazon Review Data). Standard MMRec baselines designed for clean metadata are often ill-suited for the signal ambiguity inherent in reviews. Therefore, we compare against methods established in review-enhanced and collaborative filtering contexts to ensure a fair and rigorous evaluation of semantic alignment capabilities.

\subsubsection{Implementation Details}
To ensure fair and reproducible comparisons, the baseline models were implemented within the RecBole framework \citep{Zhao2021RecBole}. The proposed SymCERE model, its related models are implemented in non-RecBole implementations, which are available anonymously on Github. For the text encoder, we used the pre-trained LLM "sarashina2.2-0.5b-instruct" for the semi-public travel dataset and "blair-roberta-base" for other datasets, using LoRA. All experiments were performed on an NVIDIA H100 GPU (90GB).

\paragraph{Fairness in Multi-Modal Comparison.}
To ensure a rigorously fair comparison, all multi-modal baselines (SMORE) utilise the exact same semantic embeddings generated by our text encoder from the user reviews. We deliberately do not use clean product metadata for baselines. This setup isolates the model architecture as the sole variable, testing each model's capability to extract signals from noisy, unstructured user feedback ("Signal Ambiguity").

\paragraph{SMORE Configuration.}
To ensure fairness, SMORE utilized the same LLM backbones. Since SMORE requires item-level features, we adapted the data using an ``Encode-Early'' strategy: training reviews for item $v$ were individually encoded (CLS token) and aggregated via mean pooling. This process strictly adhered to the temporal constraints in Section~\ref{sec:temporal_leakage} to prevent information leakage.

\paragraph{Temporal Leakage Prevention.}
\label{sec:temporal_leakage}
To prevent future information leaks, we enforce strict temporal constraints: For interaction $(u, v, t)$ at timestamp $t$, review text $r_{u,v}$ is the single review by $u$ for $v$ at $t$. We do \textit{not} aggregate reviews or use reviews after $t$. All splits respect temporal ordering (80\% earliest for training, 20\% latest for testing).

\section{Results}
This section presents the empirical results of our experiments, addressing the primary research question (RQ1) concerning the performance of our proposed SymCERE method against state-of-the-art baselines.

\subsection{Overall Performance Comparison (RQ1)}
The comprehensive performance comparison across all 15 datasets is presented in Table \ref{tab:combined_results_hr_ndcg_reordered}. The results show that our proposed multi-modal approach, particularly when implemented with an NGCF backbone, generally outperforms the suite of standard baselines across the majority of datasets and on both evaluation metrics.

\begin{table*}[t]
\centering
\caption{Overall performance comparison (HR@10 and NDCG@10) against baselines. Best result is \textbf{bolded}; second-best is \underline{underlined}. Max.\ Improv.\ (\%) is calculated against the best performing baseline.}
\label{tab:combined_results_hr_ndcg_reordered}
\resizebox{\textwidth}{!}{%
\begin{tabular}{@{}l l|cccc|c|cc|c@{}}
\toprule
\multirow{2}{*}{\textbf{Dataset}} & \multirow{2}{*}{\textbf{Metric}} & \multicolumn{4}{c|}{\textbf{GNN Baselines}} & \multicolumn{1}{c|}{\textbf{Multi-Modal}} & \multicolumn{2}{c|}{\textbf{Ours (\modelname)}} & \multirow{2}{*}{\textbf{\begin{tabular}[c]{@{}c@{}}Max.\ \\ Improv.\ (\%)\end{tabular}}} \\
\cmidrule(l){3-7} \cmidrule(l){8-9}
& & \textbf{NeuMF} & \textbf{SGL} & \textbf{NGCF} & \textbf{LightGCN} & \textbf{SMORE} & \textbf{+ NGCF} & \textbf{+ LightGCN} & \\
\midrule
\multirow{2}{*}{Clothing, Shoes \& Jewelry} & HR@10 & 0.521 & 0.530 & 0.525 & 0.530 & 0.434 & \underline{0.538} & \textbf{0.550} & $\uparrow 3.8\%$ \\
& NDCG@10 & 0.415 & 0.412 & \underline{0.420} & 0.411 & 0.311 & \textbf{0.425} & 0.398 & $\uparrow 1.2\%$ \\
\midrule
\multirow{2}{*}{CDs \& Vinyl} & HR@10 & 0.453 & 0.456 & 0.450 & 0.419 & 0.322 & \underline{0.488} & \textbf{0.495} & $\uparrow 8.5\%$ \\
& NDCG@10 & \underline{0.324} & 0.309 & \underline{0.324} & 0.285 & 0.218 & \textbf{0.332} & 0.308 & $\uparrow 2.6\%$ \\
\midrule
\multirow{2}{*}{Tools \& Home Improvement} & HR@10 & 0.330 & 0.336 & 0.323 & 0.334 & 0.181 & \underline{0.402} & \textbf{0.412} & $\uparrow 22.5\%$ \\
& NDCG@10 & 0.170 & \underline{0.175} & 0.166 & 0.174 & 0.096 & \textbf{0.187} & 0.162 & $\uparrow 6.7\%$ \\
\midrule
\multirow{2}{*}{Sports \& Outdoors} & HR@10 & 0.318 & 0.376 & 0.364 & 0.337 & 0.218 & \underline{0.424} & \textbf{0.434} & $\uparrow 14.0\%$ \\
& NDCG@10 & \underline{0.209} & 0.206 & 0.202 & 0.191 & 0.121 & \textbf{0.226} & 0.195 & $\uparrow 8.1\%$ \\
\midrule
\multirow{2}{*}{Beauty \& Personal Care} & HR@10 & 0.312 & 0.312 & 0.311 & 0.283 & 0.197 & \underline{0.386} & \textbf{0.405} & $\uparrow 29.8\%$ \\
& NDCG@10 & \underline{0.141} & 0.135 & 0.138 & 0.124 & 0.078 & \textbf{0.162} & 0.135 & $\uparrow 15.1\%$ \\
\midrule
\multirow{2}{*}{Automotive} & HR@10 & 0.408 & 0.441 & 0.424 & 0.434 & 0.312 & \underline{0.446} & \textbf{0.470} & $\uparrow 6.6\%$ \\
& NDCG@10 & 0.260 & \underline{0.281} & 0.278 & 0.276 & 0.195 & \textbf{0.285} & 0.265 & $\uparrow 1.4\%$ \\
\midrule
\multirow{2}{*}{Toys \& Games} & HR@10 & 0.312 & 0.329 & 0.312 & 0.332 & 0.233 & \underline{0.415} & \textbf{0.419} & $\uparrow 26.2\%$ \\
& NDCG@10 & 0.170 & 0.174 & \underline{0.175} & 0.173 & 0.115 & \textbf{0.198} & 0.173 & $\uparrow 13.0\%$ \\
\midrule
\multirow{2}{*}{Pet Supplies} & HR@10 & 0.268 & 0.279 & 0.268 & 0.234 & 0.174 & \underline{0.373} & \textbf{0.387} & $\uparrow 38.7\%$ \\
& NDCG@10 & \underline{0.104} & 0.101 & 0.101 & 0.086 & 0.050 & \textbf{0.129} & 0.102 & $\uparrow 24.2\%$ \\
\midrule
\multirow{2}{*}{Health \& Household} & HR@10 & \underline{0.256} & \underline{0.256} & 0.246 & 0.254 & 0.193 & \textbf{0.371} & 0.360 & $\uparrow 44.9\%$ \\
& NDCG@10 & 0.106 & \underline{0.108} & 0.102 & 0.106 & 0.063 & \textbf{0.132} & 0.107 & $\uparrow 22.2\%$ \\
\midrule
\multirow{2}{*}{Video Games} & HR@10 & 0.227 & 0.255 & 0.226 & 0.254 & 0.162 & \textbf{0.345} & \underline{0.343} & $\uparrow 35.3\%$ \\
& NDCG@10 & 0.099 & \underline{0.109} & 0.098 & 0.104 & 0.063 & \textbf{0.132} & \underline{0.109} & $\uparrow 21.1\%$ \\
\midrule
\multirow{2}{*}{Cell Phones \& Accessories} & HR@10 & 0.254 & \underline{0.306} & 0.296 & 0.299 & 0.296 & \textbf{0.411} & \textbf{0.411} & $\uparrow 34.5\%$ \\
& NDCG@10 & 0.114 & 0.146 & \underline{0.152} & 0.140 & 0.120 & \textbf{0.180} & 0.151 & $\uparrow 18.3\%$ \\
\midrule
\multirow{2}{*}{Patio, Lawn \& Garden} & HR@10 & 0.216 & 0.358 & 0.351 & 0.354 & \underline{0.372} & 0.433 & \textbf{0.450} & $\uparrow 25.8\%$ \\
& NDCG@10 & 0.130 & \underline{0.195} & \underline{0.195} & 0.192 & 0.161 & \textbf{0.225} & 0.189 & $\uparrow 15.4\%$ \\
\midrule
\multirow{2}{*}{Office Products} & HR@10 & 0.195 & 0.351 & 0.347 & 0.351 & \underline{0.355} & \textbf{0.430} & \textbf{0.430} & $\uparrow 22.4\%$ \\
& NDCG@10 & 0.093 & \underline{0.183} & 0.181 & 0.178 & 0.132 & \textbf{0.190} & 0.161 & $\uparrow 3.9\%$ \\
\midrule
\multirow{2}{*}{Yelp} & HR@10 & 0.238 & 0.193 & 0.182 & 0.146 & 0.074 & \textbf{0.344} & \underline{0.285} & $\uparrow 34.4\%$ \\
& NDCG@10 & 0.078 & 0.054 & 0.052 & 0.037 & 0.032 & \textbf{0.112} & \underline{0.082} & $\uparrow 43.6\%$ \\
\midrule
\multirow{2}{*}{Travel (JP)} & HR@10 & 0.768 & 0.742 & 0.737 & 0.695 & - & \underline{0.768} & \textbf{0.824} & $\uparrow 7.3\%$ \\
& NDCG@10 & 0.268 & \underline{0.296} & 0.269 & 0.279 & - & 0.269 & \textbf{0.380} & $\uparrow 28.4\%$ \\
\bottomrule
\end{tabular}%
}
\end{table*}

\section{Discussion}
\label{sec:discussion}
The experiments detailed in the previous section show that our proposed method, SymCERE, outperforms a range of strong baselines. These performance gains raise a question: what are the underlying mechanisms that enable this performance? In this section, we deconstruct our model's behavior to answer this question. We first revisit our geometric hypotheses (RQ2) before conducting a series of qualitative case studies to understand the semantic nature of the learned alignment and its impact on performance across different domains (RQ3).

\subsection{A Geometric View of Debiasing and Performance Gains (RQ2)}
\label{sec:ablation}
Our methodology is built on the hypothesis that mitigating popularity bias and sampling noise can be achieved through geometric controls over the embedding space. In this section, we quantitatively demonstrate the validity of this hypothesis using empirical data.

First, we posited a "magnitude-as-noise" hypothesis. Our ablation studies (Table \ref{tab:ablation_lightgcn_final_corrected_v2}) show that a model variant without L2 normalisation performs poorly. This is particularly evident with the non-linear NGCF backbone, where performance drops to near-zero levels. This confirms that without spherical projection, popularity bias (encoded as magnitude) dominates the embedding space, drowning out the semantic directional signals. L2 normalization is thus a structural prerequisite.

\begin{figure}[t]
    \centering
    \includegraphics[width=\linewidth]{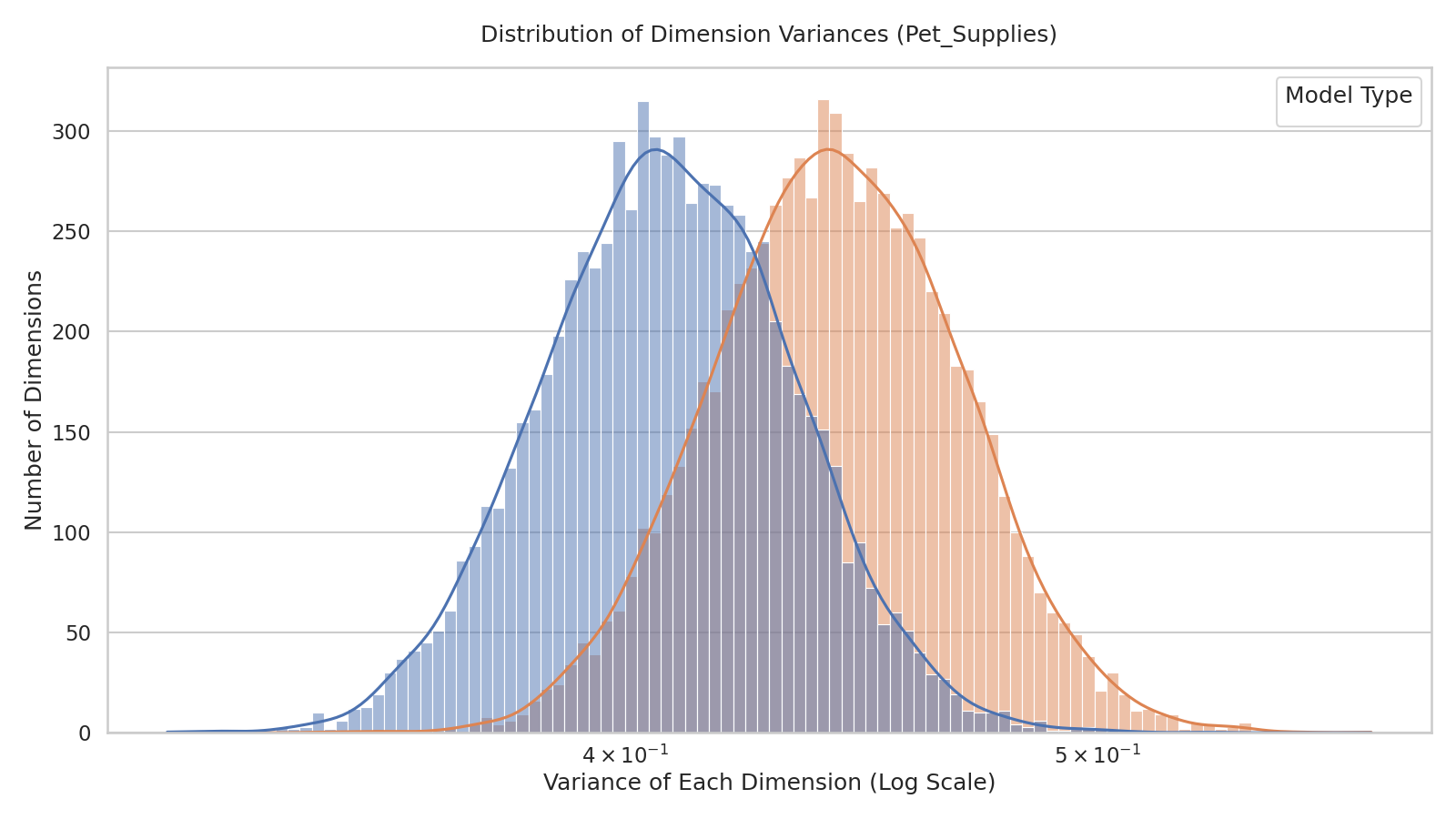}
    \caption{Impact of L2 Normalization on Dimensional Utilization (Pet Supplies). The orange distribution (With Normalization) exhibits significantly higher variance across dimensions compared to the blue distribution (Without Normalization). This confirms that SymCERE prevents dimensional collapse, utilizing the embedding space more uniformly to encode semantic differences.}
    \label{fig:dimension_variance}
\end{figure}

L2 normalisation addresses this issue and provides the foundation for improving the effectiveness of contrastive learning. We theorized that normalisation improves Uniformity by spreading embeddings more evenly across the unit hypersphere to maximize representational capacity. This claim is empirically supported by the data in Table~\ref{tab:uniformity_stats} and visualized in Figure \ref{fig:dimension_variance}. Across all datasets, applying L2 normalisation increased the standard deviation of cosine similarities and the variance of dimensions, providing quantitative and visual evidence that the embeddings are more uniformly distributed, thus forming a more discriminative representation space.

\begin{table*}[t]
\centering
\caption{Analysis of Fusion Mechanisms (Ablation Study). \texttt{MM w/ InfoNCE} represents the standard CL fusion baseline (i.e., without false negative removal). \texttt{w/o Norm} removes L2 normalization. "Min. Drop" indicates the percentage drop in performance relative to the full model (SymCERE w/ Norm).}
\label{tab:ablation_lightgcn_final_corrected_v2}
\setlength{\tabcolsep}{3pt}
\resizebox{\textwidth}{!}{%
\begin{tabular}{@{}l l | rrrr | rrrr @{}}
\toprule
\multirow{2}{*}{\textbf{Dataset}} & \multirow{2}{*}{\textbf{Metric}} & \multicolumn{4}{c|}{\textbf{NGCF Backbone}} & \multicolumn{4}{c}{\textbf{LightGCN Backbone}} \\
\cmidrule(lr){3-6} \cmidrule(lr){7-10}
& & 
\makecell[c]{\textbf{SymCERE} \\ \textbf{w/o Norm}} & \makecell[c]{\textbf{InfoNCE} \\ \textbf{w/ Norm}} & \makecell[c]{\textbf{SymCERE} \\ \textbf{w/ Norm}} & \makecell[c]{\textbf{Min.\ Drop} \\ \textbf{(\%)}} & \makecell[c]{\textbf{SymCERE} \\ \textbf{w/o Norm}} & \makecell[c]{\textbf{InfoNCE} \\ \textbf{w/ Norm}} & \makecell[c]{\textbf{SymCERE} \\ \textbf{w/ Norm}} & \makecell[c]{\textbf{Min.\ Drop} \\ \textbf{(\%)}}
\\
\midrule
\multirow{2}{*}{Clothing, Shoes \& Jewelry} & HR@10 & 0.002 & 0.536 & 0.538& $\downarrow 99.6\%$ & 0.548 & 0.547 & 0.550 & $\downarrow 0.6\%$ \\
& NDCG@10 & 0.001 & 0.428 & 0.425 & $\downarrow 99.8\%$ & 0.385 & 0.384 & 0.398 & $\downarrow 3.6\%$ \\
\midrule
\multirow{2}{*}{CDs \& Vinyl} & HR@10 & 0.007 & 0.490 & 0.488 & $\downarrow 98.6\%$ & 0.492 & 0.463 & 0.495 & $\downarrow 6.4\%$ \\
& NDCG@10 & 0.002 & 0.338 & 0.332 & $\downarrow 99.4\%$ & 0.303 & 0.295 & 0.308 & $\downarrow 4.2\%$ \\
\midrule
\multirow{2}{*}{Tools \& Home Imp.} & HR@10 & 0.005 & 0.397 & 0.402 & $\downarrow 98.8\%$ & 0.409 & 0.396 & 0.412 & $\downarrow 3.8\%$ \\
& NDCG@10 & 0.003 & 0.186 & 0.187 & $\downarrow 98.4\%$ & 0.156 & 0.153 & 0.162 & $\downarrow 6.0\%$ \\
\midrule
\multirow{2}{*}{Sports \& Outdoors} & HR@10 & 0.002 & 0.426 & 0.424 & $\downarrow 99.5\%$ & 0.433 & 0.422 & 0.434 & $\downarrow 2.9\%$ \\
& NDCG@10 & 0.001 & 0.227 & 0.226 & $\downarrow 99.6\%$ & 0.187 & 0.184 & 0.195 & $\downarrow 5.6\%$ \\
\midrule
\multirow{2}{*}{Beauty \& Personal Care} & HR@10 & 0.001 & 0.390 & 0.386 & $\downarrow 99.7\%$ & 0.401 & 0.374 & 0.405 & $\downarrow 7.7\%$ \\
& NDCG@10 & 0.000 & 0.158 & 0.162 & $\downarrow 100\%$ & 0.128 & 0.122 & 0.135 & $\downarrow 9.6\%$ \\
\midrule
\multirow{2}{*}{Automotive} & HR@10 & 0.005 & 0.450 & 0.446 & $\downarrow 98.9\%$ & 0.469 & 0.459 & 0.470 & $\downarrow 2.3\%$ \\
& NDCG@10 & 0.002 & 0.283 & 0.285 & $\downarrow 99.3\%$ & 0.254 & 0.252 & 0.265 & $\downarrow 4.9\%$ \\
\midrule
\multirow{2}{*}{Toys \& Games} & HR@10 & 0.001 & 0.417 & 0.415 & $\downarrow 99.8\%$ & 0.416 & 0.407 & 0.419 & $\downarrow 3.0\%$ \\
& NDCG@10 & 0.000 & 0.197 & 0.198 & $\downarrow 100\%$ & 0.164 & 0.162 & 0.173 & $\downarrow 6.3\%$ \\
\midrule
\multirow{2}{*}{Pet Supplies} & HR@10 & 0.004 & 0.371 & 0.373 & $\downarrow 98.9\%$ & 0.383 & 0.369 & 0.387 & $\downarrow 4.6\%$ \\
& NDCG@10 & 0.001 & 0.124 & 0.129 & $\downarrow 99.2\%$ & 0.096 & 0.093 & 0.102 & $\downarrow 8.9\%$ \\
\midrule
\multirow{2}{*}{Health and Household} & HR@10 & 0.002 & 0.355 & 0.358 & $\downarrow 99.4\%$ & 0.343 & 0.291 & 0.357 & $\downarrow 18.5\%$ \\
& NDCG@10 & 0.001 & 0.130 & 0.131 & $\downarrow 99.2\%$ & 0.102 & 0.098 & 0.112 & $\downarrow 12.5\%$ \\
\midrule
\multirow{2}{*}{Video Games} & HR@10 & 0.006 & 0.331 & 0.332 & $\downarrow 98.2\%$ & 0.280 & 0.222 & 0.326 & $\downarrow 31.9\%$ \\
& NDCG@10 & 0.002 & 0.129 & 0.130 & $\downarrow 98.5\%$ & 0.099 & 0.089 & 0.116 & $\downarrow 23.3\%$ \\
\midrule
\multirow{2}{*}{Cell Phones and Acc.} & HR@10 & 0.005 & 0.397 & 0.408 & $\downarrow 98.8\%$ & 0.408 & 0.357 & 0.401 & $\downarrow 11.0\%$ \\
& NDCG@10 & 0.001 & 0.170 & 0.178 & $\downarrow 99.4\%$ & 0.146 & 0.149 & 0.161 & $\downarrow 7.5\%$ \\
\midrule
\multirow{2}{*}{Patio, Lawn and Garden} & HR@10 & 0.007 & 0.440 & 0.433 & $\downarrow 98.4\%$ & 0.440 & 0.447 & 0.450 & $\downarrow 2.2\%$ \\
& NDCG@10 & 0.002 & 0.217 & 0.225 & $\downarrow 99.1\%$ & 0.179 & 0.182 & 0.189 & $\downarrow 5.3\%$ \\
\midrule
\multirow{2}{*}{Office Products} & HR@10 & 0.004 & 0.430 & 0.430 & $\downarrow 99.1\%$ & 0.430 & 0.422 & 0.430 & $\downarrow 1.8\%$ \\
& NDCG@10 & 0.002 & 0.203 & 0.190 & $\downarrow 98.9\%$ & 0.147 & 0.147 & 0.161 & $\downarrow 8.8\%$ \\
\midrule
\multirow{2}{*}{Yelp} & HR@10 & 0.017 & 0.350 & 0.344 & $\downarrow 95.1\%$ & 0.224 & 0.210 & 0.293 & $\downarrow 28.3\%$ \\
& NDCG@10 & 0.003 & 0.113 & 0.112 & $\downarrow 97.3\%$ & 0.070 & 0.068 & 0.089 & $\downarrow 23.6\%$ \\
\midrule
\multirow{2}{*}{Travel (JP)} & HR@10 & 0.040 & 0.789 & 0.768 & $\downarrow 94.8\%$ & 0.829 & 0.695 & 0.824 & $\downarrow 15.7\%$ \\
& NDCG@10 & 0.006 & 0.298 & 0.269 & $\downarrow 97.8\%$ & 0.390 & 0.228 & 0.380 & $\downarrow \textbf{40.0\%}$ \\
\bottomrule
\end{tabular}%
}
\end{table*}

\begin{table*}[htbp]
\centering
\caption{Detailed statistical comparison of similarity distributions, including key distributional statistics. Metrics for models with and without L2 normalisation are presented for each dataset.}
\label{tab:uniformity_stats}
\resizebox{\textwidth}{!}{%
\begin{tabular}{@{}l|cccccc|cccccc@{}}
\toprule
& \multicolumn{6}{c}{\textbf{Without Normalisation}} & \multicolumn{6}{c}{\textbf{With Normalisation}} \\
\cmidrule(l){2-7} \cmidrule(l){8-13}
\textbf{Dataset} & \textbf{Mean} & \textbf{Std. Dev.} & \textbf{Min} & \textbf{25\%} & \textbf{75\%} & \textbf{Max} & \textbf{Mean} & \textbf{Std. Dev.} & \textbf{Min} & \textbf{25\%} & \textbf{75\%} & \textbf{Max} \\
\midrule
Pet Supplies & 0.001 & 0.006 & -0.037 & -0.003 & 0.005 & 0.028 & 0.001 & $\uparrow$ 0.010 & -0.040 & -0.006 & 0.008 & $\uparrow$ 0.046 \\
Office Products & 0.000 & 0.007 & -0.030 & -0.003 & 0.006 & 0.026 & 0.000 & $\uparrow$ 0.011 & -0.037 & -0.007 & 0.008 & $\uparrow$ 0.040 \\
Clothing, Shoes \& Jewelry & 0.001 & 0.008 & -0.034 & -0.004 & 0.007 & 0.035 & 0.001 & $\uparrow$ 0.011 & -0.038 & -0.006 & 0.008 & $\uparrow$ 0.048 \\
\bottomrule
\end{tabular}%
}
\end{table*}

\subsection{Qualitative Case Studies: From Semantics to Performance (RQ3)}
\label{sec:case_study}
The quantitative results and the geometric intuition above raise a deeper qualitative question: what semantic information does the model actually use to align embeddings, and why does its effectiveness vary across datasets? To answer this (RQ3), we performed in-depth case studies on three representative datasets exhibiting high, medium, and low performance gains, using LIME \citep{Ribeiro2016Why} to analyze influential review terms.

\subsubsection{High-Improvement Case: Pet Supplies}
The \textit{Pet Supplies} dataset showed one of the largest performance improvements (e.g., +38.7\% in HR@10). Our analysis reveals that the model's success stems from its ability to prioritize objective, information-rich words. The nature of these words shifts from technical specifications to product functionalities and target subjects.

As summarized in Table \ref{tab:lime_analysis_pet_supplies}, the model identifies keywords related to pet type (`puppy`, `cat`), product category (`crate`, `leash`, `toy`), and functional attributes (`chew`, `durable`, `healthy`) as strong positive contributors to similarity.

Consider a review like: “My puppy loves this durable chew toy. Perfect size for teething.” The model learns that objective terms like `puppy`, `durable`, and `chew toy` provide a stronger alignment signal than the generic sentiment `loves`. These keywords are information-dense and describe the specific context of use, allowing the model to distinguish between a toy for a teething puppy and a food bowl for a large dog. This semantic precision explains the substantial performance lift. The model resolves "Signal Ambiguity" by focusing on the reason for the user's interaction, not just the interaction itself.

\subsubsection{Mid-Improvement Case: Office Products}
The \textit{Office Products} dataset represents a middle-ground case. A key characteristic of this domain is its blend of reviews containing objective, factual terms with those describing subjective user experience. For instance, in reviews concerning printer ink, objective terms that directly indicate product compatibility or malfunction—such as ‘HP’, ‘genuine’, ‘clogged’, and ‘dry’—are correctly identified by the model as important signals. On the other hand, reviews for writing instruments like fountain pens are centered on subjective terms dependent on personal taste, such as ‘nib’, ‘writing feel’, and ‘design’.

We postulate that this duality is what leads to the moderate performance improvement. The model likely achieves high precision in fact-based sub-domains, such as printer-related issues, but is partially confounded by signal ambiguity in the highly subjective sub-domain of writing instruments. Consequently, the overall rate of improvement is not as pronounced as in a more uniformly objective dataset like Pet Supplies.

\subsubsection{Low-Improvement Case: Clothing, Shoes and Jewelry}
In contrast, the \textit{Clothing, Shoes and Jewelry} dataset saw less improvement. Our analysis suggests this is due to a combination of factors that create high signal ambiguity.

Firstly, the domain is dominated by subjective aesthetic judgements (‘beautiful’, ‘style’, ‘elegant’) that lack an objective benchmark. Secondly, even seemingly factual terms like ‘size’ and ‘fit’ are personal and context-dependent. Thirdly, our LIME analysis revealed a presence of non-English (primarily Spanish) reviews (e.g., ‘producto’, ‘calidad’). If the underlying language model is mainly English-trained, this can act as noise, affecting alignment accuracy. This combination of subjectivity, contextual dependency, and multilingual noise (summarized in Table \ref{tab:lime_analysis_clothing}) can make it difficult for the model to extract a clear semantic signal, thereby limiting performance gains.

\begin{table*}[htbp]
\centering
\caption{LIME analysis summary of word contributions to similarity in the Pet Supplies dataset. The model prioritizes objective terms related to product function, type, and attributes.}
\label{tab:lime_analysis_pet_supplies}
\resizebox{\linewidth}{!}{%
\begin{tabular}{@{}lccc@{}}
\toprule
\textbf{Keyword Type} & \textbf{Example Words} & \textbf{Contribution} & \textbf{Reason} \\
\midrule
Product Function/Type & `chew`, `toy`, `crate`, `leash` & Strong Signal & Objective, high-impact functional term \\
Target/Attribute & `puppy`, `dog`, `durable`, `healthy` & Supporting Signal & Objective attribute of product/user \\
Generic Sentiment & `love`, `happy`, `great`, `nice` & Weak / Ambiguous & Subjective, lacks specific details \\
Stop-words & `I`, `my`, `the`, `a`, `it` & Noise / Detrimental & High frequency, no semantic value \\
\bottomrule
\end{tabular}}
\end{table*}

\begin{table*}[htbp]
\centering
\caption{LIME analysis summary of word contributions to similarity in the Office Products dataset. The model leverages a mix of objective product specifications and subjective feedback.}
\label{tab:lime_analysis_office_products_short}
\resizebox{\linewidth}{!}{%
\begin{tabular}{@{}lccc@{}}
\toprule
\textbf{Keyword Type} & \textbf{Example Words} & \textbf{Contribution} & \textbf{Reason} \\
\midrule
Product Spec/Function & `ink`, `printer`, `cartridge`, `genuine` & Strong Signal & Objective, high-impact spec/function \\
Subjective Experience & `nib`, `pen`, `smooth`, `paper` & Context-Dependent Signal & Subjective user experience/taste \\
Generic Sentiment & `great`, `good`, `nice`, `works` & Weak / Ambiguous & Subjective, lacks specific details \\
Stop-words & `I`, `my`, `the`, `a`, `it` & Noise / Detrimental & High frequency, no semantic value \\
\bottomrule
\end{tabular}}
\end{table*}


\begin{table*}[htbp]
\centering
\caption{LIME analysis summary for the Clothing, Shoes and Jewelry dataset. The model may struggle with highly subjective, context-dependent, and multilingual terms.}
\label{tab:lime_analysis_clothing}
\resizebox{\linewidth}{!}{%
\begin{tabular}{@{}lccc@{}}
\toprule
\textbf{Keyword Type} & \textbf{Example Words} & \textbf{Contribution} & \textbf{Reason} \\
\midrule
Product Spec/Function & `size`, `fit`, `small`, `large` & Ambiguous & Depends on user's body/preference \\
Subjective Experience & `beautiful`, `elegant`, `style`, `looks` & Highly Ambiguous & Personal taste, no objective benchmark \\
Generic Sentiment & `love`, `nice`, `good`, `wear` & Weak / Ambiguous & Low information, lacks specific reason \\
Foreign Language & `producto`, `calidad`, `excelente` & Noise / Mismatch & Potential language model mismatch \\
\bottomrule
\end{tabular}}
\end{table*}

\begin{table*}[htbp]
\centering
\caption{Examples of review texts providing Strong/Objective versus Weak/Subjective signals. Objective product attributes (\underline{underlined}) serve as strong anchors for alignment, while subjective or anecdotal terms (\underline{underlined}) provide weaker signals.}
\label{tab:review_examples_new_fullwidth}
\begin{tabularx}{\textwidth}{@{}llX@{}}
\toprule
\textbf{Dataset} & \textbf{Signal Type} & \textbf{Review Text with Highlighted Keywords} \\
\midrule
\multirow{4}{*}{\makecell[l]{Pet \\ Supplies}} %
& \multirow{2}{*}{\makecell[l]{Strong / \\ Objective}} & i rescued a toy poodle and she has \underline{no teeth}. i feed her canned foods because of the \underline{softness} and was feeding ... \\
\cmidrule(l){3-3}
& & This is a great container, and perfectly holds a \underline{30 pound bag} of Taste of the Wild dog food. The wheels move ...\\
\cmidrule(l){2-3}
& \multirow{2}{*}{\makecell[l]{Weak / \\ Subjective}} & We had this for about 3 months before our puppy realized he could \underline{tear it apart}. We had this on the bottom ... \\
\cmidrule(l){3-3}
& & My one year old cat... took one of the packs of treats, opened it (\underline{ripped it open}) and I caught her red-pawed ... \\
\midrule
\multirow{4}{*}{\makecell[l]{Office \\ Products}} %
& \multirow{2}{*}{\makecell[l]{Strong / \\ Objective}} & Stick with the original, generic inks will \underline{damage your machine}. The \underline{print quality} is better with HP and the ... \\
\cmidrule(l){3-3}
& & The \underline{ink dried up} in the lines ruining my machine. Also, the print quality was not as good. Keep away. \\
\cmidrule(l){2-3}
& \multirow{2}{*}{\makecell[l]{Weak / \\ Subjective}} & I ordered this pen based on its \underline{looks}, the Parker \underline{reputation} and its reasonable price. The nib that came with it ... \\
\cmidrule(l){3-3}
& & I have always come to expect \underline{great quality} from Parker products. This is no exception. I am very satisfied ... \\
\midrule
\multirow{4}{*}{\makecell[l]{Clothing, \\ Shoes \\ \& Jewelry}} %
& \multirow{2}{*}{\makecell[l]{Strong / \\ Objective}} & I really do love this hat. It is \underline{very warm} and completely \underline{covers my ears}. I use Watch Caps to hike in cold ... \\
\cmidrule(l){3-3}
& & it wears like a verry expensive piece time piece it also feels \underline{verry light} like a feather easy to set even a child ... \\
\cmidrule(l){2-3}
& \multirow{2}{*}{\makecell[l]{Weak / \\ Subjective}} & The first set I ordered were \underline{damaged} the material had \underline{holes in it}. I returned the first set and they sent me a ... \\
\cmidrule(l){3-3}
& & I was impressed with this piece of jewelry. It was not only \underline{beautiful in appearance} but also very well made ... \\
\bottomrule
\end{tabularx}%
\end{table*}

\subsection{Summary of Findings}
Our analysis suggests that robust cross-modal alignment is a form of semantic and contextual understanding. The case studies reveal that the performance of our method is correlated with the nature of the language in a given domain. The model excels when it can ground its understanding in objective, factual, and information-rich keywords that describe a product's function, attributes, and context of use. As the language becomes more subjective and ambiguous, the model's ability to disambiguate signals and create a well-structured geometric space may diminish, limiting performance gains. This answers RQ3 and highlights that a key aspect of our model's performance is its learned ability to identify and prioritize reliable and objective signals within user-generated text. This finding implies that future work on review-based systems may benefit from focusing on objective feature extraction rather than advanced sentiment analysis.

\section{Limitations and Future Work}
\label{sec:future_work}

Our study demonstrates that SymCERE yields substantial performance gains, particularly when employing a linear graph encoder like LightGCN. This success stems from the natural compatibility between LightGCN's linear propagation, which primarily encodes information in the direction of embeddings, and our methodology's reliance on L2 normalisation.

However, a potential limitation and a promising avenue for future research emerges when considering non-linear encoders such as NGCF. Our ablation study revealed that applying SymCERE to NGCF, while still effective, operates within a potential tension: NGCF's non-linear transformations may learn to encode meaningful information within the magnitude of the embedding vectors, whereas our L2 normalisation purposefully discards this magnitude to mitigate popularity bias. This suggests that the full potential of sophisticated non-linear architectures may not be realised due to this architectural conflict.

Future work could proceed in two main directions. First, one could design novel non-linear GNN architectures that are inherently compatible with hyperspherical embeddings, learning representations where directional information is paramount by design. Second, it remains an open question whether a more advanced normalisation technique could be developed---one that can disentangle and preserve potentially useful information encoded in vector magnitudes while still neutralising the detrimental effects of popularity bias. Such research would further enhance the synergy between advanced graph encoders and robust contrastive learning methods.

And while our LIME-based analysis provides compelling evidence for "semantic anchoring," we acknowledge that LIME is a local surrogate model. Therefore, a potential limitation is that our interpretation is based on local explanations, and the extent to which this mechanism represents the model's global behavior remains an open question. Future work could employ global explanation techniques to validate whether this semantic anchoring phenomenon holds true across the entire representation space.

\section{Conclusion}

We addressed the interconnected challenges of false negatives, popularity bias, and signal ambiguity with SymCERE, a unified contrastive learning method. By integrating a symmetric NCE loss with hyperspherical projection (L2 normalisation), SymCERE significantly outperforms strong baselines on both NDCG@10 and HR@10 across 15 datasets (e.g., +43.6\% NDCG@10). Our key finding, termed ‘semantic anchoring’, reveals the model learns to prioritize objective, information-rich, and domain-specific attributes over superficial sentiment, showing that robust alignment stems from factual understanding. This unified and principled approach of structurally mitigating biases thus creates a path toward recommender systems that are not only more accurate but also more robust, interpretable, and fair by promoting item diversity.

\begin{acks}
This work was supported by JSPS KAKENHI Grant Numbers JP23K28908, JP25K21210.
\end{acks}

\bibliographystyle{ACM-Reference-Format}
\bibliography{main}

\end{document}